\magnification=\magstep1
\baselineskip=18pt
\centerline{\bf ELEVENTH-ORDER CALCULATION OF GREEN'S FUNCTIONS IN THE}
\centerline{\bf ISING LIMIT FOR ARBITRARY SPACE-TIME DIMENSION $\bf D$}
\bigskip
\bigskip
\bigskip
\centerline{Carl M. Bender}
\centerline{Department of Physics}
\centerline{Washington University}
\centerline{St. Louis, MO 63130}
\medskip
\centerline{and}
\medskip
\centerline{Stefan Boettcher}
\centerline{Department of Physics}
\centerline{Brookhaven National Laboratory}
\centerline{Upton, NY 11973}
\bigskip
\bigskip
\bigskip
\bigskip
\centerline{\bf ABSTRACT}
\bigskip
This paper extends an earlier high-temperature lattice calculation of the
renormalized Green's functions of a $D$-dimensional Euclidean scalar quantum
field theory in the Ising limit. The previous calculation included all graphs 
through sixth order. Here, we present the results of an eleventh-order 
calculation. The extrapolation to the continuum limit in the previous 
calculation was rather clumsy and did not appear to converge when $D>2$. Here, 
we present an improved extrapolation which gives uniformly good results for all 
real values of the dimension between $D=0$ and $D=4$. We find that the 
four-point Green's function has the value $0.62 \pm 0.007$ when $D=2$ and 
$0.98 \pm 0.01$ when $D=3$ and that the six-point Green's function has the value
$0.96 \pm 0.03$ when $D=2$ and $1.2 \pm 0.2$ when $D=3$.
\footnote{}{PACS numbers: 11.10.Ef, 11.10.Jj, 11.10.Np}
\vfill \eject

\baselineskip= 23pt
There have been many attempts to calculate the coefficients in the effective
potential of a Euclidean scalar quantum field theory in the Ising limit. These
coefficients are just the dimensionless renormalized $2n$-point Green's
functions evaluated at zero momentum. Techniques that have been used to
determine these Green's functions include high-temperature lattice expansions,
Monte Carlo methods, and epsilon expansions. For the case $D=3$ a complete list
of references is given in a recent paper by Tsypin.$^{1}$

In a series of papers$^{2,3}$ high-temperature lattice techniques were used to
obtain the dependence of the Green's functions upon the Euclidean space-time
dimension $D$ for $D$ ranging continuously between 0 and 4. In this work a
strong-coupling calculation to sixth order was performed analytically on a
hypercubic lattice in $D$ dimensions and Pad\'e extrapolation techniques were
invented to obtain the continuum limit.$^{4}$ The current paper is an addendum
to the work in Refs.~2 and 3. Here, we extend the strong-coupling calculation to
eleventh order. Furthermore, we use an improved Pad\'e extrapolation method that
relies on information taken from the results of a large-$N$ calculation and our
recent studies of dimensional expansions for quantum field theory.$^{5,6,7}$

Our strong-coupling lattice calculations are identical to those described in
Ref.~2 except that the graphs were generated using a FORTRAN program and
evaluated analytically using MACSYMA. The eleventh-order calculation involves 
several hundred times as many graphs as the sixth-order calculation. We have
verified the accuracy of our expansions for the specific cases of $D=2$, $D=3$,
and $D=4$ dimensions by comparing them with previous calculations$^{8}$.

The lattice series for the renormalized four-point and six-point Green's
functions are as follows:
\vfill\eject
\baselineskip=12pt
$$\eqalign{
\gamma_4={y^{-D/2}\over 12} &\bigm [1+4\,D\,y +(4\,D^{2}-10\,D)\,y^2 +16\,D\,y^3
+(30 \,D-80 \,D^{2}) \,y^4 \cr
\noalign{\medskip}
&+(256\,D^{3}+104\,D^{2}-192\,D)\,y^5\cr
\noalign{\medskip}
&+(-704\,D^{4}-1736\,D^{3}+2508\,D^{2}-656\,D)\,y^6 \cr
\noalign{\medskip}
&+(1792\,D^{5}+10432\,D^{4}-11232\,D^{3}-3872\,D^{2}+4992\,D)\,y^7 \cr
\noalign{\medskip}
&+(-4352\,D^{6}-45600\,D^{5}+11456\,D^{4}+123672\,D^{3}\cr
&~~-128440\,D^{2}+ 35542\,D)\,y^8 \cr
\noalign{\medskip}
&+(10240\,D^{7}+168320\,D^{6}+181248\,D^{5}-1052576\,D^{4}\cr
&~~+ 2615584/3 ~ D^{3}+ 76664\,D^{2}- 681472/3 ~D) \,y^9 \cr
\noalign{\medskip}
&+(-23552\,D^{8}-558208\,D^{7}-1630272\,D^{6}+5391904\,D^{5} \cr
&~~-1011536/3~D^{4}-10102936\,D^{3}+29622092/3~D^{2} -2720752\,D) \,y^{10}\cr
\noalign{\medskip}
&+(53248\,D^{9}+1718272\,D^{8}+9081856\,D^{7}-18274816\,D^{6}\cr
&~~- 113682176/3 ~D^{5} + 367432576/3 ~ D^{4} - 292976128/3 ~D^{3} \cr
&~~+ 18425408/3 ~ D^{2} + 14757984\,D)\,y^{11} + \ldots \bigm ]} \eqno{(1)}$$
and
$$\eqalign{
\gamma_6={y^{-D}\over 30} &\bigm [ 1 +6\,D \,y +(12\,D^{2}-6\,D)\,y^2
+(8\,D^{3}-12\,D^{2}-20\,D)\,y^3 \cr
\noalign{\medskip}
&+(48\,D^{2}+48\,D)\,y^4 +(-96\,D^{3}-816\,D^{2}+528\,D)\,y^5\cr
\noalign{\medskip}
&+(192\,D^{4}+4640\,D^{3}-2736\,D^{2}-560\,D)\,y^6 \cr
\noalign{\medskip}
&+(-384\,D^{5}-18432\,D^{4}-10800\,D^{3}+46512\,D^{2}-23040\,D)\,y^7 \cr
\noalign{\medskip}
&+(768\,D^{6}+61440\,D^{5}+188352\,D^{4}-510816\,D^{3}+357324\,D^{2}-
72492\,D)\,y^8 \cr
\noalign{\medskip}
&+(-1536\,D^{7}-184576\,D^{6}-1274880\,D^{5}+2653440\,D^{4}\cr
&~~-77496\,D^{3}- 2911496\,D^{2}+1698240\,D)\,y^9 \cr
\noalign{\medskip}
&+(3072\,D^{8}+517632\,D^{7}+6280704\,D^{6}-6584832\,D^{5}-27745840\,D^{4}\cr
&~~+65401176\,D^{3}-49332608\,D^{2}+11853912\,D)\,y^{10} \cr
\noalign{\medskip}
&+(-6144\,D^{9}-1382400\,D^{8}-25928448\,D^{7}-13343232\,D^{6}+286690784\,D^{5}
\cr
&~~ -516057392\,D^{4}+211594432\,D^{3}
+210150872\,D^{2}-153291336\, D)\,y^{11} + \ldots \bigm ] , }\eqno{(2)}$$
\vfill\eject
\baselineskip=23pt
\noindent
where $y=(Ma)^{-2}$, $a$ is the lattice spacing, and $M$ is the renormalized
mass, which is obtained from the two-point function as explained in Ref.~2.

The quantity $\sqrt{y}$ is the dimensionless correlation length. The continuum
limit $a\to 0$ corresponds to infinite correlation length. To obtain the
continuum Green's functions it is necessary to extrapolate the formulas in (1)
and (2) to their values at $y=\infty$. Direct extrapolation to the continuum 
limit of either series in (1) or (2) leads to a sequence of extrapolants that 
becomes badly behaved when $D$ increases beyond 2; we find that extrapolations 
as functions of $D$ do not converge to a limiting curve (see Figs.~1 and 2).
However, for $D$ near 0 these extrapolants {\sl are} well behaved and converge
rapidly to the known exact values$^{2}$ $\gamma_4 = 1/4$ and $\gamma_6 = 1/4$ at
$D=1$.

To improve our extrapolation we make the following observation. We consider a
scalar field theory having an $O(N)$ symmetry. The model we have studied above
corresponds to the case $N=1$. In the limit $N\to\infty$ one can solve for the
Green's functions exactly. We obtain the following lattice results:

$$N \gamma_4 ^{(N=\infty)} = {y^{-D/2} \over 4 \int_0^{\infty} dt\,t
e^{-t} [e^{-2ty}I_0 (2ty)]^D }\eqno{(3)}$$
and
$$N^2 \gamma_6 ^{(N=\infty)} = {y^{-D} \int_0^{\infty} dt\,t^2 e^{-t} [e^{-2ty}
I_0 (2ty)]^D \over 12 \left ( \int_0^{\infty} dt\,t\,e^{-t} [e^{-2ty}I_0 (2ty)]^D
\right ) ^3 },\eqno{(4)}$$
where we have summed over the external indices. In the continuum limit
$y\to\infty$ we have
$$N \gamma_4 ^{(N=\infty)} = {(4\pi)^{D/2} \over 4~\Gamma (2-D/2)} \eqno{(5)}$$
and
$$N^2 \gamma_6 ^{(N=\infty)} = {(4\pi)^{D} \Gamma(3-D/2) \over
12~[\Gamma (2-D/2)]^3},\eqno{(6)}$$
where $D$ lies in the range $0\leq D\leq 4$.
Each of these functions rises from its value at $D=0$, attains a maximum, and
falls to 0 at $D=4$.

Under the assumption that the Green's functions for $N=1$ vanish at $D=4$ like
those in (5) and (6) we can extract such a behavior from the series (1) and (2)
by performing a Borel summation as follows. Consider the lattice series in (1)
for $\gamma_4$. This series has the general form
$$\gamma_4 ={y^{-D/2}\over 12} \sum_{k=0}^{\infty} P_k (D) y^k ,\eqno{(7)}$$
where $P_k (D)$ are polynomials of maximum degree $k$. One can read off the
first eleven polynomials $P_k (D)$ from (1). We can rewrite (7) as
$${1\over 12 \gamma_4} = y^{D/2} \sum_{k=0}^{\infty} Q_k (D) y^k,\eqno{(8)}$$
where $Q_k (D)$ is another polynomial in $D$. Next, we insert the identity
$$1= {1\over (k+1)!} \int_0^{\infty} dt\,t^{k+1} e^{-t} \eqno{(9)}$$
for each term in the sum. This converts (8) to the form
$${1\over 12\gamma_4} = \int_0^{\infty} dt\,t^{1-D/2} e^{-t} f(yt),\eqno{(10)}$$
where we define
$$\eqalign{
f(x) &= x^{D/2} \sum_{k=0}^{\infty} {Q_k (D) x^k\over (k+1)!}\cr
	 &= \left (x \sum_{k=0}^{\infty} R_k(D) x^k\right )^{D/2},}\eqno{(11)}$$
where again $R_k(D)$ are polynomials in $D$.

We now take the continuum limit of the expression (10). Assuming that
$f(\infty)$ exists in the limit $y\to\infty$ so that (10) separates into a
product of two terms, we have
$$\gamma_4 = {1\over 12\,\Gamma(2-D/2)\,f(\infty)}.\eqno{(12)}$$
Thus, we have forced the continuum limit of the four-point function to take the
form of the large-$N$ result in (5) apart from $f(\infty)$, which is a smoothly
varying function of $D$.

There is an immediate indication that the Borel summation leading to (12) has
a significant impact. We find that the polynomials $R_k(D)$ in (11) are
significantly simpler than the original polynomials $P_k (D)$ in (7); the
polynomials $R_k (D)$ have maximum degree $[k/2] - 1$, about half the degree of
the polynomials $P_k (D)$. The polynomials $R_k (D)$, $k=0,~1,~2,~ \ldots,11$,
are as follows:

{\baselineskip=12pt
$$\eqalign{
R_{0} (D)&=1,\cr
\noalign{\medskip}
R_{1} (D)&=-2,\cr
\noalign{\medskip}
R_{2} (D)&=11/3,\cr
\noalign{\medskip}
R_{3} (D)&=-16/3,\cr
\noalign{\medskip}
R_{4} (D)&= - {7\over 45}  \,D+  {233\over 36} ,\cr
\noalign{\medskip}
R_{5} (D)&= {2\over 3} \,D- {125\over 18} ,\cr
\noalign{\medskip}
R_{6} (D)&= {73\over 2835} \,D^2- {1346\over 945} \,D+ {76691\over 11340},\cr
\noalign{\medskip}
R_{7} (D)&= - {221\over 1134} \,D^2 + {2713\over 1260} \,D-{16939\over 2835},\cr
\noalign{\medskip}
R_{8} (D)&= - {13\over 8100} \,D^3+  {98323\over 170100} \,D^2
- {8837\over 3240} \,D+ {2627137\over 544320} ,\cr
\noalign{\medskip}
R_{9} (D)&= {163\over 3150} \,D^3- {14699\over 14175} \,D^2+
 {17111\over 5670} \,D- {235681\over 64800} ,\cr
\noalign{\medskip}
R_{10} (D)&=- {457\over 267300} \,D^4- {1338257\over 
5613300} \,D^3+ {2015579\over 1403325} \,D^2- {9257497\over 
3207600} \,D+ {230357209\over 89812800} ,\cr
\noalign{\medskip}
R_{11} (D)&=- {611\over 69300} \,D^4+ {6066953\over 11226600} \,D^3
- {3923681\over 2245320} \,D^2+ {110092603\over 
44906400} \,D- {75132389\over 44906400} .
}\eqno{(13)}$$
}
\medskip
The resummation of the lattice series as performed above reduces the problem of
extracting the continuum limit to finding the value of $f(\infty)$. This is done
using the same Pad\'e techniques as were used in Ref.~2. If we perform this
numerical calculation we obtain a sequence of approximants, one for each new
order in perturbation theory. The first eleven such approximants for $\gamma_4$
in (12) are plotted in Fig.~3. Each approximant is a continuous function of $D$
for $0\leq D\leq 4$. Note that the approximants are smooth and well behaved; the
sequence is monotone increasing and appears to converge uniformly to a limiting
curve. The dotted line on Fig.~3 is an extrapolation of these approximants to
this limiting curve obtained using Richardson extrapolation.$^{9}$

There are a number of ways to assess the accuracy of the limiting curve. First, 
one can Taylor expand this limiting curve about $D=0$ as a series in powers of
$D$. This Taylor series has the form
$$\gamma_4^{\rm limiting~curve}(D)={1\over 12} (1+1.18\,D+0.64\,D^2+0.19\,D^3+
0.03\,D^4 +\ldots).\eqno{(14)}$$
We may then compare this Taylor series with that recently obtained$^{7}$ using
dimensional expansion methods:
$$\gamma_4^{\rm dimensional~expansion}(D)={1\over 12}(1+1.18\,D+0.62\,D^2+0.18\,
D^3+ 0.03\,D^4 +\ldots).\eqno{(15)}$$
Note that the coefficients of these two series are almost identical.
Second, we can examine the limiting curve at $D=1$, for which the exact
value $\gamma_4=1/4$ is known. At this value of $D$ the limiting curve
has the value $0.2526$ so it is slightly high by about 1\%. 

In Figs.~4 and 5 we demonstrate how we obtain the limiting curve for the cases
$D=2$ and $D=3$. We have plotted the $n$th-order Richardson extrapolants 
for the approximants in Fig.~3 for $n=1,~2,~\ldots,~5$ versus the inverse order 
of the approximants. We then determine where each of these extrapolants crosses 
the vertical axis (each intersection is indicated by a horizontal bar). Finally,
we extrapolate to the limiting value of these intersection points. This 
procedure gives the value $\gamma_4=0.620\pm 0.007$ at $D=2$, indicated in
Fig.~4 by a fancy square. This result is to be compared with $\gamma_4=0.6108\pm
0.0025$ obtained by Baker and Kincaid$^{10}$. Similarly, in Fig.~5 we find that 
the limiting curve gives $\gamma_4=0.986\pm 0.010$ at $D=3$. This value compares
reasonably well with previous results, as tabulated in Ref.~1. For example, 
Baker and Kincaid$^{11}$ obtain $0.98$, Monte Carlo studies give results between
$0.9$ and $1$, and renormalization group studies give results around $0.98$.
Note that the limiting curve in Fig.~3 has a maximum extremely close to $D=3$; 
numerically, the maximum occurs at $D=3.03$. 

The same procedure that was used to extrapolate (1) to the continuum and thereby
to obtain a plot of $\gamma_4$ as a function of $D$ can be applied to (2). We
perform a Borel summation of the series in (2) as follows. The lattice series in
(2) for $\gamma_6$ has the general form
$$\gamma_6 = {y^{-D}\over 30} \sum_{k=0}^{\infty} S_k (D) y^k ,\eqno{(16)}$$
where $S_k (D)$ are polynomials of maximum degree $k$. One can read off the
first eleven polynomials $S_k (D)$ from (2). From the structure of
$\gamma_6 ^{(N=\infty)}$ in (4) we are motivated to rewrite (16) in the form
$${y^{D/2}\int_0^{\infty} dt\,t^2 e^{-t} [e^{-2ty}I_0(2ty)]^D\over 30\gamma_6}
= \left [ y^{D/2} \sum_{k=0}^{\infty} T_k (D) y^k \right ]^3,\eqno{(17)}$$
where $T_k (D)$ is another polynomial in $D$. Again, we insert the identity (9)
for each term in the sum in (17). This converts (17) to the form

$${y^{D/2}\int_0^{\infty} dt\,t^2 e^{-t} [e^{-2ty}I_0(2ty)]^D\over 30\gamma_6}
=\left [ \int_0^{\infty} dt\,t^{1-D/2} e^{-t} g(ty)\right ]^3,\eqno{(18)}$$
where we define
$$\eqalign{
g(x) &= x^{D/2} \sum_{k=0}^{\infty} {T_k (D) x^k\over (k+1)!}\cr
	 &= \left (x \sum_{k=0}^{\infty} U_k(D) x^k\right )^{D/2},}\eqno{(19)}$$
where $U_k(D)$ are polynomials in $D$ of degree $[k/2] -1$ similar in
structure to those in (13).

Next, we take the continuum limit of the expression (18). Assuming that
$g(\infty)$ exists in the limit $y\to\infty$ we find that (18) separates into a
product of several terms and we have
$$\gamma_6 = {(4\pi)^{-D/2} \Gamma(3-D/2)\over 30 [\Gamma(2-D/2)\,g(\infty)]^3}.
\eqno{(20)}$$
Thus, we have forced the continuum limit of the six-point function to take the
form of the large-$N$ result given in (6) apart from $g(\infty)$, which is a
function of $D$.

Again, the resummation of the lattice series reduces the problem of extracting
the continuum limit to finding the value of $g(\infty)$. This is done using the
same Pad\'e techniques as were used in Ref.~2. We perform this numerical
calculation and obtain a sequence of approximants, one for each new order in
perturbation theory. The first eleven such approximants for $\gamma_6$ in (20)
are plotted in Fig.~6. Each approximant is a continuous function of $D$ for
$0\leq D \leq 4$. As in Fig.~3 the approximants are smooth and well behaved; the
sequence is monotone increasing and appears to converge uniformly to a limiting
curve indicated in Fig.~6 by a dotted line. This limiting curve is again
obtained using fifth-order Richardson extrapolation. The limiting curve at $D=1$
passes through the value $\gamma_6=0.240$ which differs from the exact value
$\gamma_6=1/4$ by about 4\%.

The limiting curve predicts that $\gamma_6=0.96\pm 0.04$ at $D=2$ and $\gamma_6=
1.2\pm 0.1$ at $D=3$. This value is lower than most previous results, as
tabulated in Ref.~1, but it is certainly larger than zero. By comparison, an
epsilon expansion around $D=4$ gives$^{1,12}$
$${\gamma_6 \over (\gamma_4)^2} = 2\epsilon -{20\over
27}\epsilon^2+1.2759\,\epsilon^3 + \ldots~.\eqno{(21)}$$
This series appears to be divergent but a direct optimal truncation of the
series after one term with $\epsilon=1$ gives the value $\gamma_6=1.9\pm 0.7$.
(Here, we have substituted the value $\gamma_4 = 0.986$ given above.) However, 
if we perform a $(1,1)$-Pad\'e summation of this series, which seems justified 
because of the alternating sign pattern, we obtain the smaller value 
$\gamma_6=1.66\pm 0.28$, in better agreement with our predicted value.

 Finally, we observe that the maxima of $\gamma_{2i}$ as a function of $D$
appear to follow a pattern. We observed already that $\gamma_4$ has a maximum
that is close to $D=3$. Here, we find that the limiting curve for $\gamma_6$ has
a maximum at $D=2.66$ which is very close to $8/3$. An interesting conjecture is
that in general the maximum might be located at $D_{max}={2(i+1)\over i}$, the
value of $D$ for which a $\phi^{2i+2}$ theory becomes free.
\bigskip
\bigskip

We thank the U. S. Department of Energy for funding this research.
\vfill \eject

\centerline{\bf REFERENCES}
\bigskip
\item{$^{1}$} M. M. Tsypin, Aachen preprint PITHA 94/9 and hep-lat/9401034
and references therein.
\medskip
\item{$^2$} C. M. Bender, F. Cooper, G. S. Guralnik, R. Roskies, and D. H.
Sharp, Phys. Rev. Lett. {\bf 43}, 537 (1979).
\medskip
\item{$^3$} C. M. Bender, F. Cooper, G. S. Guralnik, and D. H. Sharp, Phys. Rev.
D {\bf 19}, 1865 (1979).
\medskip
\item{$^4$} C. M. Bender, Los Alamos Science {\bf 2}, 76 (1981). See also
R. J. Rivers, Phys. Rev. D {\bf 20}, 3425 (1979) and Phys. Rev. D {\bf 22}, 3135
(1980).
\medskip
\item{$^{5}$} C. M. Bender, S. Boettcher, and L. N. Lipatov, Phys. Rev. Lett.
{\bf 65}, 3674 (1992).
\medskip
\item{$^{6}$} C. M. Bender, S. Boettcher, and L. N. Lipatov, Phys. Rev. D
{\bf 46}, 5557 (1992).
\medskip
\item{$^{7}$} C. M. Bender and S. Boettcher, Phys. Rev. D {\bf 48}, 4919 (1993).
\medskip
\item{$^{8}$} M. L\"uscher and P. Weisz, Nucl. Phys. {\bf B300}, 325 (1988).
\medskip
\item{$^{9}$} C. M. Bender and S. A. Orszag, {\it Advanced Mathematical Methods
for Scientists and Engineers}, (McGraw-Hill, New York, 1978), pp.~375-376.
\medskip
\item{$^{10}$} G. A Baker in {\it Phase Transitions and Critical Phenomena},
Vol.~9, C. Domb and J. L. Lebowitz, eds. (Academic, London, 1984), pp.~233-311.
\medskip
\item{$^{11}$} G. A. Baker and J. M. Kincaid, Phys. Rev. Lett. {\bf 42}, 1431
(1979).
\medskip
\item{$^{12}$} J. Zinn-Justin, {\it Quantum Field Theory and Critical
Phenomena}, (Clarendon, Oxford, 1989), Sect.~25.1.3.
\medskip

\vfill \eject
\centerline{\bf FIGURE CAPTIONS}
\bigskip

\noindent
Figure 1. The first $k$ Pad\'e approximants, $k=1,~2,~\ldots,~11$, to the 
continuum limit of the four-point Green's function. These approximants are
constructed from the lattice series in (1) using the Pad\'e procedure explained
in Ref.~2. Observe that the approximants are well behaved when $D$ is near 0; in
particular, they converge nicely to the exact value 1/4 at $D=1$ (indicated by a
plus sign). However, when $D$ increases beyond 2 the approximants become
irregular and do not seem to converge to a limiting function of $D$. (Some of
the approximants reach zero and terminate as $D$ increases because they become
complex.)
\medskip

\noindent
Figure 2. Same as in Fig.~1 except that we have plotted the first $k$
approximants to the six-point Green's function for $k=1,~2,~\ldots,~11$. Again,
we find that the approximants converge to the exact value 1/4 at $D=1$ but that 
they are irregular for $D>2$.
\medskip

\noindent
Figure 3. First eleven approximants to $\gamma_4$ plotted as functions of $D$
for $0\leq D\leq4$ [see (12)]. The approximants form a monotone increasing
sequence of curves; the labeling indicates the order. Note that the approximants
are smooth curves that seem to be tending uniformly to a limiting curve. We have
obtained this limiting curve (dotted curve) by means of fifth-order Richardson
extrapolation$^{9}$. The exact result $\gamma_4 = 1/4$ at $D=1$ is indicated by
a plus sign; the limiting curve passes within 1\% of this point.
\medskip

\noindent
Figure 4. Plot of the $n$th-order Richardson extrapolants for the approximants
in Fig.~3 for $n=1,~2,~\ldots,~5$ versus the inverse order of the approximants 
for the case $D=2$. Each Richardson extrapolant is linearly extended 
(dash-dotted line) until it intersects the vertical axis. Each intersection is 
indicated by a horizontal bar. We then extrapolate to the limiting value of
these intersection points, indicated by a fancy square. This procedure predicts
that $\gamma_4=0.620$ at $D=2$.
\medskip

\noindent
Figure 5. Same as in Fig.~4 except that $D=3$. This extrapolation procedure
predicts that $\gamma_4=0.986$ at $D=3$.
\medskip

\noindent
Figure 6. First eleven approximants to $\gamma_6$ plotted as functions of $D$
for $0\leq D\leq4$ [see (20)]. The approximants form a monotone increasing
sequence of curve as indicated by the labeling. As in Fig.~3 for the four-point
function, the approximants are smooth curves that seem to be tending uniformly
to a limiting curve. This limiting curve (dotted curve) is a fifth-order
Richardson extrapolation. The exact result $\gamma_6 =1/4$ at $D=1$ is indicated
by a plus sign; the limiting curve passes within 4\% of this point.
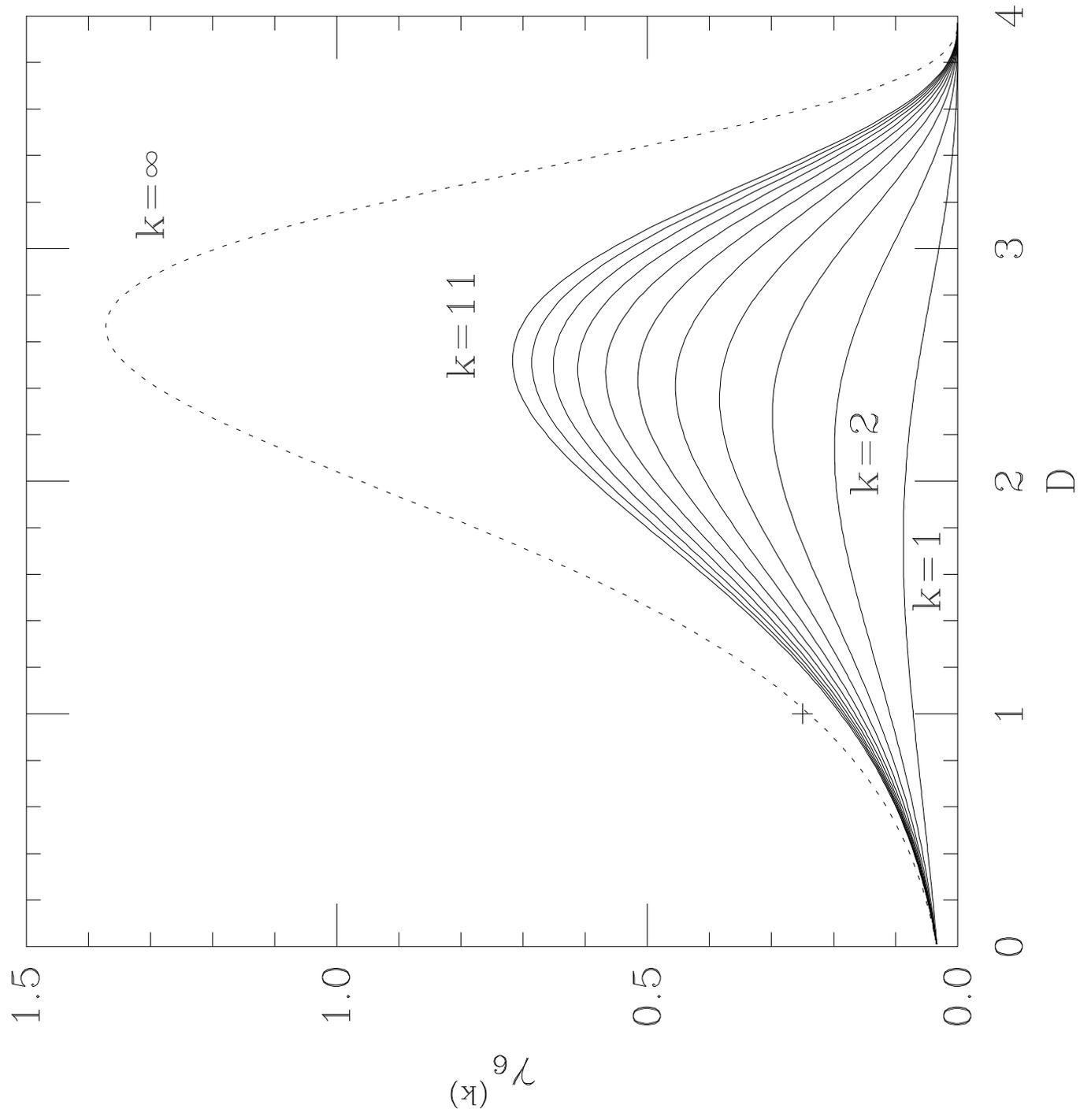
\bye